\documentclass[sigconf]{acmart}
\usepackage[colorinlistoftodos]{todonotes}
\usepackage{verbatim}
\usepackage{tabularx}
\usepackage{soul}
\usepackage{enumitem}
\usepackage{booktabs}

\usepackage{blindtext}
\usepackage[font=small,skip=3pt]{caption}
\copyrightyear{2020} 
\acmYear{2020} 
\setcopyright{acmlicensed}\acmConference[SIGIR '20]{Proceedings of the 43rd International ACM SIGIR Conference on Research and Development in Information Retrieval}{July 25--30, 2020}{Virtual Event, China}
\acmBooktitle{Proceedings of the 43rd International ACM SIGIR Conference on Research and Development in Information Retrieval (SIGIR '20), July 25--30, 2020, Virtual Event, China}
\acmPrice{15.00}
\acmDOI{10.1145/3397271.3401189}
\acmISBN{978-1-4503-8016-4/20/07}

\graphicspath{ {figures/} }

\begin{document}
\fancyhead{}

%
% The "title" command has an optional parameter, allowing the author to define a "short title" to be used in page headers.
%\title{Product Insights: Using web search to better understand products}
\title{Studying Ransomware Attacks Using Web Search Logs}

%
% The "author" command and its associated commands are used to define the authors and their affiliations.
% Of note is the shared affiliation of the first two authors, and the "authornote" and "authornotemark" commands
% used to denote shared contribution to the research.

\author{Chetan Bansal*}
\thanks{* These authors contributed equally to the work.}
\email{chetanb@microsoft.com}
\affiliation{%
  \institution{Microsoft Research}
  \streetaddress{}
  \city{Redmond}
  \state{WA, USA}
}

\author{Pantazis Deligiannis*}
\email{pdeligia@microsoft.com}
\affiliation{%
  \institution{Microsoft Research}
  \streetaddress{}
  \city{Redmond}
  \state{WA, USA}
}

\author{Chandra Maddila*}
\email{chmaddil@microsoft.com}
\affiliation{%
  \institution{Microsoft Research}
  \streetaddress{}
  \city{Redmond}
  \state{WA, USA}
}

\author{Nikitha Rao*}
\email{t-nirao@microsoft.com}
\affiliation{%
  \institution{Microsoft Research}
  \streetaddress{}
  \city{Bangalore}
  \state{India}
}

%
% By default, the full list of authors will be used in the page headers. Often, this list is too long, and will overlap
% other information printed in the page headers. This command allows the author to define a more concise list
% of authors' names for this purpose.
\renewcommand{\shortauthors}{Chetan Bansal, Pantazis Deligiannis, Chandra Maddila, and Nikitha Rao}

%
% The abstract is a short summary of the work to be presented in the article.
\begin{abstract}
Cyber attacks are increasingly becoming prevalent and causing significant damage to individuals, businesses and even countries. In particular, ransomware attacks have grown significantly over the last decade. We do the first study on mining insights about ransomware attacks by analyzing query logs from Bing web search engine. We first extract ransomware related queries and then build a machine learning model to identify queries where users are seeking support for ransomware attacks. We show that user search behavior and characteristics are correlated with ransomware attacks. We also analyse trends in the temporal and geographical space and validate our findings against publicly available information. Lastly, we do a case study on `Nemty', a popular ransomware, to show that it is possible to derive accurate insights about cyber attacks by query log analysis.
\end{abstract}
% \footnote{https://www.mcafee.com/enterprise/en-us/assets/reports/rp-quarterly-threats-aug-2019.pdf}

\begin{comment}
\begin{CCSXML}
<ccs2012>
<concept>
<concept_id>10002978.10003022.10003026</concept_id>
<concept_desc>Security and privacy~Web application security</concept_desc>
<concept_significance>500</concept_significance>
</concept>
<concept>
<concept_id>10002951.10003317.10003325.10003328</concept_id>
<concept_desc>Information systems~Query log analysis</concept_desc>
<concept_significance>500</concept_significance>
</concept>
<concept>
<concept_id>10002951.10003260.10003261.10003263</concept_id>
<concept_desc>Information systems~Web search engines</concept_desc>
<concept_significance>300</concept_significance>
</concept>
</ccs2012>
\end{CCSXML}

\ccsdesc[500]{Security and privacy~Web application security}
\ccsdesc[500]{Information systems~Query log analysis}
\ccsdesc[300]{Information systems~Web search engines}
\end{comment}


%
% This command processes the author and affiliation and title information and builds
% the first part of the formatted document.
\maketitle
\section{Introduction}
With internet becoming ubiquitous, cyber attacks have become increasingly prevalent. Cyber attacks have had significant negative implications in major sectors such as healthcare, finance, manufacturing, etc. As per a study by the Internet Society\cite{isoc-report}, over 5 billion data records were exposed in 2018 and the total cost of these attacks was estimated to be over \$45 billion. There are several different types of attacks such as phishing, wiretapping, denial of service, ransomware, etc. In particular, with the advent of cryptocurrencies, which are used for making payments to attackers, ransomware attacks have increased exponentially over the last decade. Ransomware encrypts and locks the victims' files until they pay money to the attackers. The financial damage due to ransomware attacks is estimated to increase from \$8 billion in 2018 to \$20 billion in 2021 \cite{isoc-report}. Several cities and organizations have been impacted by these attacks \cite{nytimes-report} affecting millions of people in the process.

Given the significance of these attacks, it's critical to understand the impact and scale of these attacks. Most of the information available today is from manually curated public repositories \cite{kharraz2015cutting} or by cyber security vendors \cite{mcafee-report}. However, these suffer from fragmentation and delays. So, it's important to explore alternative data sources and methods to improve our understanding of these attacks. Web search query logs offer a unique way to do population-scale analysis \cite{jhaver2019measuring, bansal2019usage}. For instance, Paparrizos et al. \cite{paparrizos2016detecting} and Xu et al. \cite{xu2011predicting} have analyzed health related search queries to predict trends for various diseases and epidemics. Chancellor et al. \cite{chancellor2018measuring} have shown that macro-economic factors like employment demand can be characterized using query logs. In the security domain, Canali et al. \cite{canali2014effectiveness} used the browsing history of users to predict the risk of visiting malicious websites.

In this work, we do the first study to analyze cyber attacks, specifically ransomware, using query logs from Bing, a major web search engine. We mine ransomware related queries from anonymized query logs and use machine learning models to extract queries where users are seeking support for ransomware attacks. This is critical since we want to analyze queries and sessions where the users were likely attacked rather than those who were just looking for information about these attacks. Next, we do feature correlation analysis to understand if search behavior and user attributes are correlated with attacks. We also report on temporal and geographic trends for users who were seeking support for ransomware attacks. Lastly, we do a case study on the Nemty ransomware \cite{Nemty} and show that just by query log analysis we are able to learn about the origin and the effectiveness of the attack.
\vspace{-1mm}
\section{Data}
\label{sec:dataset}
We use the anonymized query logs from Bing to perform our analysis which is conducted over a four month time span between July 1st, 2019 and October 31st, 2019. As web search patterns tend to vary significantly based on several factors, we focus this study on queries from US region with English locale. However, the methodology is generic and can be expanded to other regions and locales.

\vspace{-1mm}
\subsection{Terminology}
Below, we define some key terms that we use throughout the paper:
\begin{enumerate}[leftmargin=3mm,itemsep=0mm]
    \item \textbf{Ransomware Queries} - Ransomware related queries having the keyword `ransomware' in the query or the clicked URL(s).
    \item \textbf{Support Queries} - Ransomware queries indicating that the user is trying to seek solutions for attacks. Sample queries: `how to recover encrypted files', `.besub ransomware decryption software'.
    \item \textbf{Non-support Queries} - Ransomware queries where the intent is not to find support or solution for attacks but for seeking general information, facts, etc. Sample queries: `top ransomware attacks', `20 Texas cities attacked with ransomware'. 
    \item \textbf{Attacked Users} - Users who searched for at least one support query for ransomware attacks.
    \item \textbf{Safe Users} - Users who did not search for any support queries for attacks. We have randomly sampled one million safe users for the study. 
\end{enumerate}
\textbf{Limitations}: Please note that since the query logs are anonymized, we lack ground truth about individual users to validate our observations. However, in Section \ref{sec:case-study}, we show that the insights from this study are consistent with public information about the attacks.

\subsection{Manual Annotation}
\label{sec:annotation}
\iffalse
Owing to the large volume of data in the query logs, manually labeling each query can be a mammoth task. For the purpose of this analysis we manually label only a small subset of the data and then train a machine learning model to find the support related queries. We randomly select an initial subset of $200$ queries such that either the query or one of the URLs clicked by the user has the keyword `ransomware' present. This is to ensure that the query is indeed related to ransomware attacks. Four annotators then individually label the $200$ samples based on the following guidelines - 
\begin{itemize}[leftmargin=3mm,itemsep=0mm]
    \item Each ransomware related query is assigned a label which is determined by looking at both the query string and the clicked URLs.
    \item The label $1$ is assigned to the query if either the query string or the clicked URL present some evidence that the user is attacked and is looking for solutions (or support) to overcome the attack. Example queries - `.access virus decryptor', `how to remove londec file by cmd', `.besub ransomware decryption software'
    \item The label $0$ is assigned if the user is not seeking any kind of support or looking for solutions after being attacked. Here the users are trying to gain some kind of information about ransomware attacks. Example queries - ` is tnt express a virus', ` Backup with Ransomware Protection free', ` 20 Texas cities attacked 2.5 million ransom'
\end{itemize}
\fi

Owing to the large volume of query logs, manually labeling each query will be a mammoth task. So, we manually label 1000 queries and then train a machine learning model to find the support queries and the attacked users. First, four annotators individually label a random sample of $200$ ransomware queries as either \textit{support queries} or \textit{non-support queries}. We then calculate the inter-annotator agreement score using Fleiss kappa \cite{fleiss1971measuring}. With the resulting score being $94.21$, translating to almost perfect agreement, each of the four annotators were asked to label a disjoint set of $200$ samples each. Including the initial set of $200$ samples, a labeled dataset of $1000$ samples was created with support queries being $32.8$\% of the data. 

\subsection{Support Query Classification}
\label{sec:classification}

The labeled data (see Section \ref{sec:annotation}) is then processed before we train a binary classification model. We tokenize the query string and the clicked URLs and compute the word embeddings of tokens that are not stopwords using a pretrained Word2Vec model \cite{word2vec}. The individual token embeddings are then aggregated together resulting in a $300$ dimension feature vector. Several classification models are trained on the data and the five-fold cross validation scores are reported in Table \ref{tab:classification_scores}.

\begin{table}[H]
\small
  \vspace{-3mm}
  \begin{center}
    \caption{Comparison of support query classifiers with 5-fold CV.}
    \label{tab:classification_scores}
    \begin{tabular}{|c|c|c|c|c|}
     \hline
      \textbf{Classifier}  & \textbf{Accuracy}  & \textbf{Precision } & \textbf{Recall } & \textbf{ F1 Score} \\
      
    \hline
    Decision Tree & 83.6 & 80 & 81 & 81 \\
    \hline      
    Random Forest & 89.1 & 94 & 70 & 80 \\
%    \hline
%    Linear SVM & 92.9 & 97 & 86 & 91\\
    \hline  
    \textbf{LinearSVC} & \textbf{93.20} & 94 & \textbf{89} & 91\\
    \hline  
    Gaussian SVM & 92.6 & \textbf{98} & 87 & \textbf{92}\\
%    \hline
%    K-Nearest Neighbors & 91.7 & 93 & 89 & 91\\
    \hline
%    Naive Bayes & 89.8 & 91 & 91 & 91\\
%    \hline
    \end{tabular}
  \end{center}
  \vspace{-3mm}
\end{table}

We observe that LinearSVC is the best performing model with the highest five-fold cross validation accuracy of $93.2$\% and a F1 score of $91$. For the $158,001$ ransomware queries that were found in the four month duration, the trained LinearSVC model was used to derive the inference labels. A total of $12,357$ unique users were identified as attacked users which corresponds to $11.64$\% of the total users that searched for ransomware queries. The resulting dataset, which is a union of all the queries searched for by the attacked users and the safe users, comprises of $23,439,988$ queries out of which $8,957,348$ queries belong to attacked users.

\section{User Behavior Analysis}

The data collected from the previous section is analysed to identify the behavioral differences in attacked users and safe users. To this end, we identify different features and group them into different categories based on the type of behaviour it indicates. The list of categories and the corresponding features are as follows:

\begin{itemize}[leftmargin=3mm,itemsep=0mm]
    \item Volume of search - number of queries, number of adult queries, dwell time, clicks, sat clicks (clicks with dwell time > $30$s \cite{fox2005evaluating}).
    \item Diversity in searches - unique URL domains.
    \item Time of search - morning ($6$AM - $7$PM), evening ($7$PM - $12$AM) or night ($12$AM - $6$AM)
    \item Day of the week - weekday (Monday to Friday) or weekend (Saturday and Sunday). 
    \item Device used - device type, operating system and browser type.
\end{itemize}

Along with the total counts, we normalize the features at a session level as well as the user level. The feature values are computed for all users in the dataset. We then analyse the differences in distribution of feature values for all attacked users and safe users. Table \ref{tab:feature_mean_comparison} summarises the percentage difference in mean values of the feature distributions of attacked users and safe users where the feature values are aggregated at a session level. Note that only the features where the percentage difference was higher than $100$\% are shown in the table.

\begin{table}[H]
\small
    \vspace{-2mm}
  \begin{center}
    \caption{Feature comparison for attacked \& safe users.}
    \label{tab:feature_mean_comparison}
    \begin{tabular}{|c|c|}
    
     \hline
      \textbf{Feature} & \textbf{Percent} \\
       & \textbf{Difference} \\
     \hline
        Total Number Of Queries & 192.16 \\
     \hline
        Total Number Of Adult Queries & 191.91 \\
     \hline
        Total Number Of Clicks & 193.25 \\
     \hline
        Total Number Of Unique URL Domains & 193.22 \\
     \hline
        Total Number Of Sat Clicks & 193.49 \\
     \hline
        Total Dwell Time & 193.55 \\
      \hline
        Total Number Of Requests At Morning & 184.22 \\
     \hline
        Total Number Of Requests At Evening & 196.72 \\
     \hline
        Total Number Of Requests At Night & 188.81 \\
     \hline
        Total Number Of Requests On Weekday & 189.21 \\
     \hline
        Total Number Of Requests On Weekend & 189.13 \\
     \hline
        Mean Total Number Of Queries & 107.84 \\
     \hline
        Mean Total Number Of Adult Queries & 103.21 \\
     \hline
        Mean Total Number Of Clicks & 110.38 \\
     \hline
        Mean Total Number Of Unique URL Domains & 110.64 \\
     \hline
        Mean Total Number Of Sat Clicks &  127.80 \\
     \hline
        Mean Total Dwell Time & 130.40 \\
     \hline
        \iffalse
        Mean Average Number Of Clicks & 8.22 \\
     \hline
        Mean Average Unique URL Domains & 8.73 \\
     \hline
        Mean Average Number Of Sat Clicks  & 51.61 \\
     \hline
        Mean Average Dwell Time & 57.11 \\
     \hline
     \fi
    \end{tabular}
  \end{center}
      \vspace{-3mm}
\end{table}

\begin{figure*}[h]
%\vspace{-2mm}
  \begin{center}
    \includegraphics[width=0.9\textwidth]{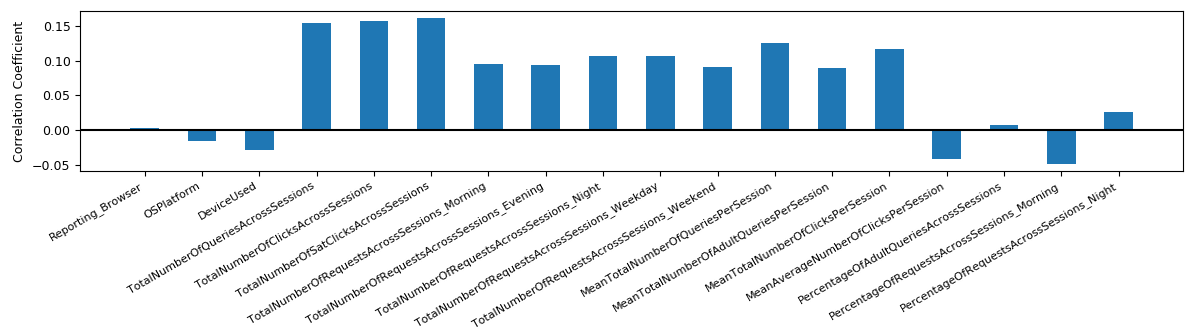}
  \end{center}
\vspace{-2mm}
\caption{Spearman's correlation coefficients between features and being attacked by ransomware.}
\label{fig:correlation_coefficient}
\vspace{-2mm}
\end{figure*}

%\subsection{Feature Correlations}
%\noindent\textbf{Feature Correlations -} 

Following the feature comparison, a feature correlation analysis is carried out using Spearman's correlation coefficient~\cite{spearmancorrelation} as it is able to capture monotonic relationships between variables without assuming the data to be of normal distribution. The values of the coefficient range from $-1$ to $1$ to denote negative and positive correlations. Once the coefficients are computed, the confidence of the results obtained is tested via a standard significance test. The correlation value of a feature is considered statistically significant if the significance level (or p-value) is less than $0.05$ indicating a confidence level of $95$\%. Figure \ref{fig:correlation_coefficient} summarizes the set of features which satisfy this threshold condition. It is evident from the coefficient values that there is very weak or no correlation between the variables and likeliness of being attacked by ransomware.

An interesting observation made was that attacked users generally had a much higher search volume compared to safe users which implies that the more the users searches the web, the more likely they are to be attacked. There was also significant positive correlation between the percentage of queries searched at night time and a negative correlation for percentage of queries searched in the morning indicating that users are more likely to get attacked at night time. Another interesting behavior seen was that attacked users had higher positive correlations with adult queries. 

\section{Trend analysis}
\subsection{Hourly Trends}
\begin{figure}[!t]
\includegraphics[width=0.75\columnwidth]{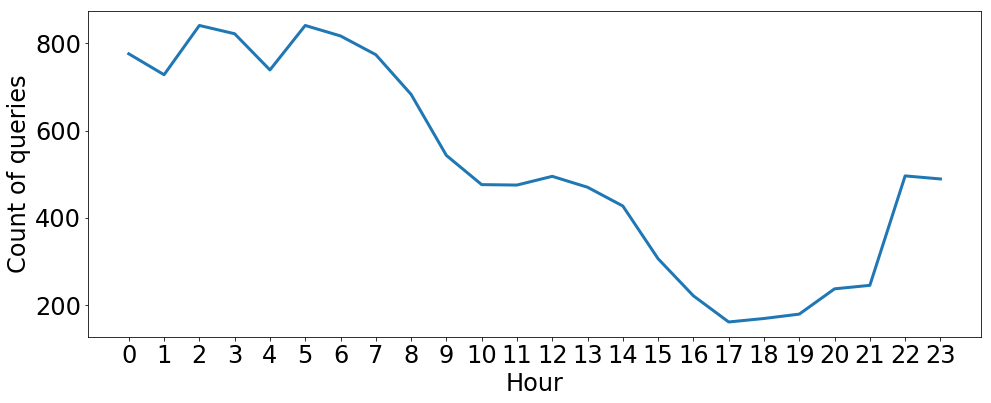}
%\vspace{-1.5\baselineskip}
\caption{Hourly distribution of support queries.}
\label{fig:HourlyTrends1}
\vspace{-4mm}
\end{figure}

\begin{figure}[!t]
\includegraphics[width=0.75\columnwidth]{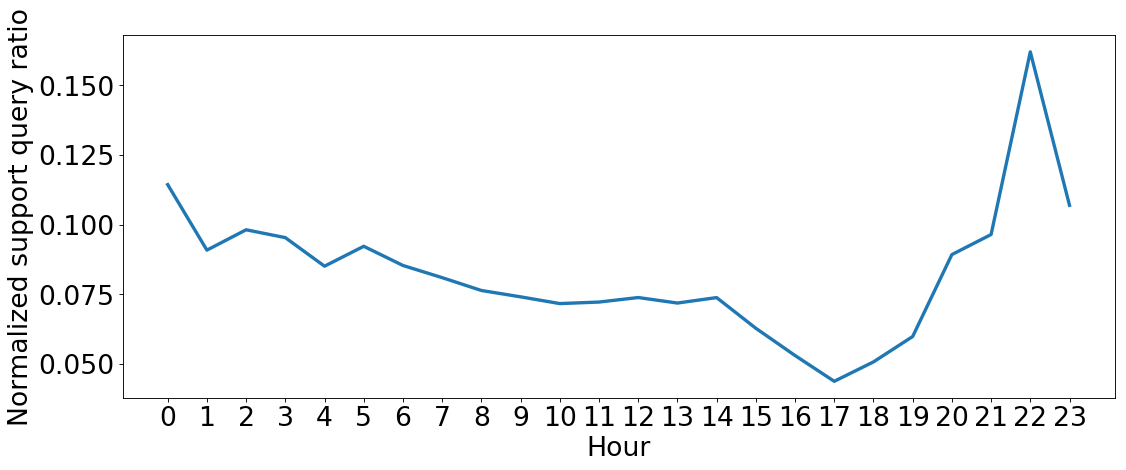}
%\vspace{-1.5\baselineskip}
\caption{Hourly distribution of normalized support query ratio.}
\label{fig:HourlyTrends2}
\vspace{-2mm}
\end{figure}

We analyzed how the behaviour of attacked users seeking solutions to ransomware attacks changes at each hour of the day by plotting hourly trends emerging from our dataset. Figure \ref{fig:HourlyTrends1}  shows that users were searching for solutions mostly during non-working hours (outside of the 9AM - 5PM window). This makes sense as users who are really determined to mitigate ransomware attacks on their own, by leveraging web search, are sparing some focused time outside of their regular working hours to find solutions. However, it could also be that the regular web search volume is very high during those hours.
To better understand the trend, we plotted a graph shown in Figure \ref{fig:HourlyTrends2} that shows the normalized distribution of ratio of support queries and how it is varying at different hours of the day. This bolsters our earlier finding (i.e., users spend more time searching for solutions outside of working hours) and highlights the fact that a lot of search activity to find solutions to ransomware attacks happens between 6PM - 11PM, which makes sense as this is the time window where users usually spare more focused time for finding solutions to their non-work problems.

\subsection{Geographical Trends}

\iffalse
\begin{figure}
\includegraphics[width=\columnwidth]{Figures/States-1.png}
\vspace{-1.5\baselineskip}
\caption{Distribution of attack related searches by state.}
\label{fig:StateTrends1}
\end{figure}
\fi

\begin{figure}[t!]
%\subfigure{ 
\begin{center}
  \includegraphics[width=0.45\columnwidth]{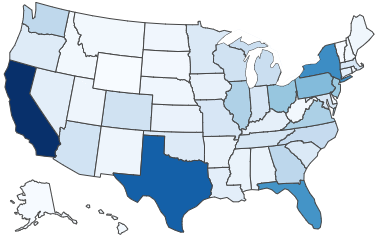}
    \includegraphics[width=0.45\columnwidth]{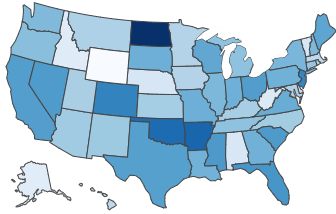}
  \end{center}
%}
\caption{Distribution of support seeking query searches by US state (absolute numbers on the left and normalized ratios on the right).}
\label{fig:state_trends}
\vspace{-5mm}
\end{figure}

In this study we focused on understanding how the search trends vary across different states in the US. In Figure \ref{fig:state_trends} (left) we plotted a heat map of how the volume of support queries vary across different states in the US. Noticeably, states that are large (in size or population or internet penetration), like California, Texas, New York, is where a lot of activity is seen. However, this could also be because the total search volume in these states is generally higher. To better understand which states record higher volumes of ransomware queries, we normalized the data by calculating the ratio of support queries compared to non-support queries in that state. This yielded some interesting insights, as seen in Figure \ref{fig:state_trends} (right): states like North Dakota, Arkansas, Oklahoma is where the ratio of support queries is high though the overall search volume is low (compared to states like California or Texas). This can also be intuitively correlated to the massive ransomware attacks that were seen in various schools and public offices in states like North Dakota and Arkansas in the year 2019  \cite{ND-Ransomware, AR-Ransomware}, which could have caused users in these states to record a higher normalized support query ratio.

\iffalse
\begin{figure}
\includegraphics[width=\columnwidth]{Figures/States-2.png}
\vspace{-1.5\baselineskip}
\caption{Distribution of attack related searches by state.}
\label{fig:StateTrends2}
\end{figure}
\fi

\section{Case Study}
\label{sec:case-study}
We present a case study to see if we can learn insights about specific attacks using query log analysis . We looked at all recent (second half of 2019) ransomware attacks with significant impact listed by the NJ Cybersecurity and Communications Integration Cell (NJCCIC)\footnote{\url{https://www.cyber.nj.gov/threat-profiles/ransomware-variants/}}. For this study, we focus on the Nemty~\cite{Nemty} ransomware, however our technique generalizes to any attack.

\begin{figure}[t!]
%\vspace{-2mm}
\includegraphics[width=0.95\columnwidth]{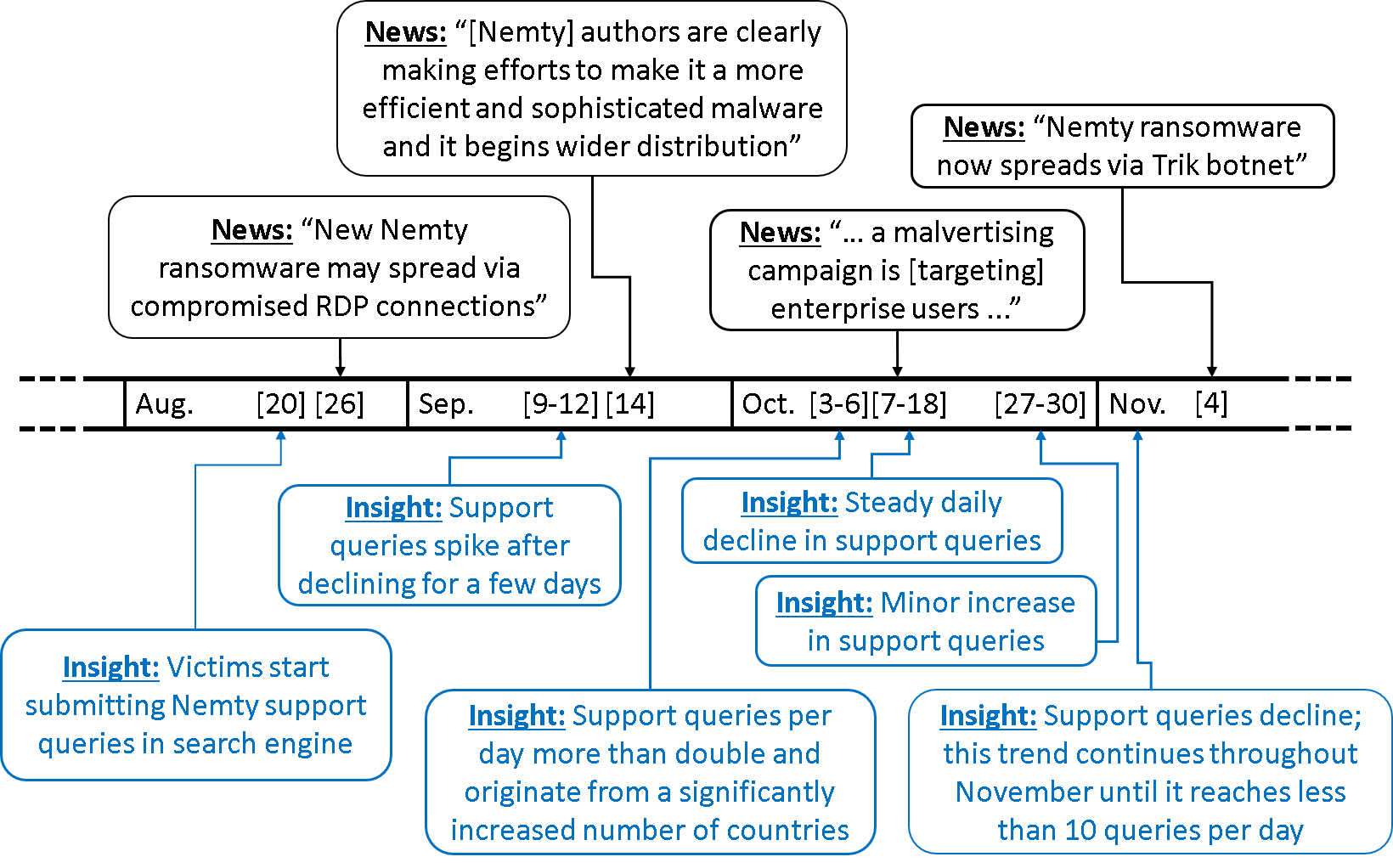}
%\vspace{-1mm}
\caption{Timeline of news and query logs insights about Nemty.}
\label{fig:case_study_timeline}
\vspace{-2mm}
\end{figure}

Nemty is a ransomware that infects Windows OS users, encrypts their files, searches and deletes any shadow copies of these files, and finally asks victims to pay a ransom for restoring their data. Nemty started affecting users end of August 2019, and spread worldwide through distribution campaigns during September, October and November 2019, as seen in the timeline of Figure~\ref{fig:case_study_timeline} (black boxes above the timeline show published news about Nemty~\cite{Nemty}).

\begin{figure}[t!]
\includegraphics[width=0.95\columnwidth]{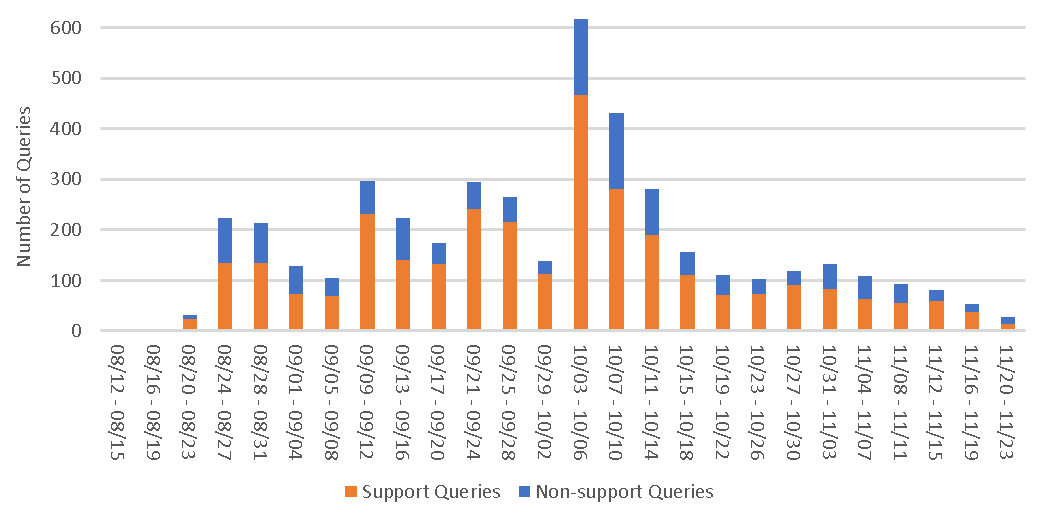}
%\vspace{-1mm}
\caption{Volume of queries about Nemty per 4-day periods.}
\label{fig:case_study}
\vspace{-2mm}
\end{figure}

We gathered attack-related search engine query logs (with English locale) about Nemty between the start of August to end of November 2019 from all countries. We then classified the queries as support and non-support using our ML model (see Section \ref{sec:classification}). Finally, we analyzed the results to gain insights about Nemty, such as when the ransomware started infecting users and how its distribution evolved over time (see blue boxes below the timeline in Figure~\ref{fig:case_study_timeline}). Our insights found trends related to the distribution of Nemty that start days before they are reported in the news~\cite{Nemty}. This result shows that our query log analysis technique could be used to timely learn about the origin and effectiveness of a ransomware attack, even in the early days of its distribution.

In Figure~\ref{fig:case_study}, we show the number of such queries per 4-day periods between August and November 2019. We see that until 08/19, there were no queries about Nemty. However, users started submitting support queries on 08/20, which is likely when the ransomware started first spreading. Indeed, on 08/26 the first news about Nemty are published~\cite{Nemty} (see Figure~\ref{fig:case_study_timeline}). On 09/14, an article~\cite{Nemty} discusses how the Nemty authors are enhancing the ransomware with the goal of achieving a wider distribution. Our analysis indeed found that although the support queries started declining early September, on 09/09 there was a spike in such queries (see Figure~\ref{fig:case_study}). The insight we gained from this corresponds to the news about the ransomware becoming more efficient and sophisticated.

There was a tremendous increase in support and non-support queries about Nemty between 10/03 and 10/06 (see Figure~\ref{fig:case_study}). Published news confirm this finding, as there was a new distribution campaign during October that was targeting enterprise users~\cite{Nemty}. We found that Nemty was now spreading worldwide, as support queries started being submitted by an increasing number of countries. After this period of time, the support queries started decreasing, most likely because more people became aware of the ransomware and how to protect from it. In early November there was a minor increase in the queries (see Figure~\ref{fig:case_study}), which corresponds to a new distribution method of Nemty via Trik botnet. However, this method was not very efficient, because after this minor increase in queries, the query volume continued to decline.

\section{Conclusion}
In this work, we did the first study to find insights about ransomware attacks using web search logs. We analyzed query logs from a major web search engine using a machine learning classifier to extract support queries by users who were attacked by ransomware. We did a correlation analysis and found that certain features such as query volume and click counts are correlated with attacks. Further, we analyzed geographical and temporal trends and validated our findings from publicly available information. Lastly, we did a case study on the Nemty ransomware and showed that with query log analysis, it is possible to mine key insights about the origin and spread of specific attacks.

%
% The next two lines define the bibliography style to be used, and the bibliography file.
\bibliographystyle{ACM-Reference-Format}
\bibliography{references}

\end{document}